\documentclass[12pt]{iopart}
\usepackage{graphicx,latexsym}

\begin{document}

\title{Multiplet Structures of Charged Fullerenes} 

\author{Ma{\l}gorzata Wierzbowska$^{1,2}$\footnote[3]{To whom correspondence should 
be addressed (wierzbom@sissa.it)}\  Martin L{\"u}ders$^{3,4}$\ Erio Tosatti$^{1,2,3}$ }  

\address{$^{1}$\ International Center for Theoretical Physics (ICTP),
Strada Costiera 11, 34100 Trieste, Italy \\
$^{2}$\ INFM DEMOCRITOS National Simulation Center, via Beirut 2--4, 34014 Trieste, Italy \\
$^{3}$\ International School for Advanced Studies (S.I.S.S.A),
Via Beirut 2-4, 34014 Trieste, Italy \\
$^{4}$\ Daresbury Laboratory, Daresbury, Warrington, WA4 4AD, United Kingdom}

\begin{abstract}
We calculated multiplet splittings for positively and negatively charged fullerene
ions within the CAS SCF method, and extracted model parameters for the intramolecular 
Hamiltonian. The method treats correctly the symmetry of ground and excited states 
for partially occupied degenerate molecular orbitals. 
We compare our results to previous calculations by the LDA, MNDO
and model SCF methods. The multiplet averaged Coulomb parameter $U$ is about 3.1 eV for 
electrons and 3.2 eV for holes.
The Hund's rule exchange parameter $J$ is found to be 113 meV 
for electrons and 192 meV for holes. 
\end{abstract}


\submitto{\JPB}

\maketitle

\newcommand{\rem}[1] {}
\newcommand{\be}{\begin{equation}}
\newcommand{\ee}{\end{equation}}
\newcommand{\bea}{\begin{eqnarray}}
\newcommand{\eea}{\end{eqnarray}}
\newcommand{\ba}{\begin{array}}
\newcommand{\ea}{\end{array}}
\newcommand{\up}{\uparrow}
\newcommand{\down}{\downarrow}
\newcommand{\Id}[1] {\int \! \! {\rm d}^3 #1}
\newcommand{\Idn}[2] {\int \! \! {\rm d}^{3 #1} #2}
\newcommand{\ID}[1] {\int \! \! \! \frac{{\rm d}^3 #1}{(2 \pi)^3}}
\newcommand{\w}{\omega}
\newcommand{\vr} {{\bf r}}
\newcommand{\vx} {{\bf x}}
\newcommand{\vy} {{\bf y}}
\newcommand{\vk} {{\bf k}}
\newcommand{\vq} {{\bf q}}
\newcommand{\vb} {{\bf b}}
\newcommand{\vR} {{\bf R}}
\newcommand{\vX} {{\bf X}}
\newcommand{\vY} {{\bf Y}}
\newcommand{\vK} {{\bf K}}
\newcommand{\vQ} {{\bf Q}}
\newcommand{\vB} {{\bf B}}
\newcommand{\vs} {{\bf s}}
\newcommand{\vS} {{\bf S}}
\newcommand{\vT} {{\bf T}}
\newcommand{\vU} {{\bf U}}
\newcommand{\vP} {{\bf P}}
\newcommand{\Y}{{\rm Y}}
\newcommand{\K}{{\rm K}}
\newcommand{\J}{{\rm J}}
\newcommand{\vG}{{\bf G}}
\newcommand{\nn} {\nonumber}
\newcommand{\eps} {\epsilon}
\newcommand{\Ave}[1] { \langle #1 \rangle }
\newcommand{\bAve}[1] { \Big\langle #1 \Big\rangle }
\newcommand{\bra}[1] { \langle #1 | }
\newcommand{\ket}[1] { | #1 \rangle }
\newcommand{\mref}[1] { (\ref{#1}) }
\newcommand{\fnd}[2] {  \DS \frac{ \delta #1 }{ \delta #2 } } 
\newcommand{\grfn}[1] 
        { \lim_{\delta \to 0^+} \int \! \! d \w' \; 
                \frac{ #1 }{ \w -\w' + i \delta} }
\newcommand{\Grfnp}[3] 
        { \frac{ #1 }{ \w + #2 + i \delta} \, #3 }
\newcommand{\Grfnm}[3] 
        { \frac{ #1 }{ \w - #2 + i \delta} \, #3 }

\renewcommand{\Im}{ {\rm Im} \, }
\newcommand{\inputEps}[2]  {   \begin{figure}  
                               \caption{#1}
                               \centerline{ \epsfxsize=10cm \epsfbox{#2} }
                               \end{figure} }
\newcounter{saveeqn}

\newcommand{\alpheqn}{\refstepcounter{equation}\setcounter{saveeqn}{\value{equation}}%
        \setcounter{equation}{0}%
        \renewcommand{\theequation}{\mbox{\arabic{chapter}.\arabic{saveeqn}-\alph{equation}}}}

\newcommand{\reseteqn}{\setcounter{equation}{\value{saveeqn}}%
        \renewcommand{\theequation}{\arabic{chapter}.\arabic{equation}}}

\newcommand{\alpheqnl}[1]{\refstepcounter{equation}\label{#1}\setcounter{saveeqn}{\value{equation}}%
        \setcounter{equation}{0}%
        \renewcommand{\theequation}{\mbox{\arabic{chapter}.\arabic{saveeqn}-\alph{equation}}}}

\newcommand{\alpheqnapp}{\refstepcounter{equation}\setcounter{saveeqn}{\value{equation}}%
        \setcounter{equation}{0}%
        \renewcommand{\theequation}{\mbox{\Alph{chapter}.\arabic{saveeqn}-\alph{equation}}}}

\newcommand{\alpheqnappl}[1]{\refstepcounter{equation}\label{#1}\setcounter{saveeqn}{\value{equation}}%
        \setcounter{equation}{0}%
        \renewcommand{\theequation}{\mbox{\Alph{chapter}.\arabic{saveeqn}-\alph{equation}}}}

\newcommand{\reseteqnapp}{\setcounter{equation}{\value{saveeqn}}%
        \renewcommand{\theequation}{\Alph{chapter}.\arabic{equation}}}

\newcommand{\header}[1] {\begin{center}
                                {\Huge {\bf #1 } }
                         \end{center}
                         \vspace*{1cm}
                        }

\section{Introduction}

Despite the fact that only very few successful attempts to produce hole-doped
crystalline fullerene have been reported (see Panich {\it et al.} 
\cite{Panich1,Panich2,Datars}), 
it is interesting to investigate the properties of such compounds. 
Their physical behavior will to large parts be determined by the ground 
state properties and the low energy excitations of positively charged molecular ions.
Besides that, fullerene molecular ions can be realized in solution, and to some extent
also in vacuum.

The physics of charged fullerenes is determined by their
high orbital degeneracy (3 in case of electron doping and 5 in case of
hole doping for C$_{60}$), and the resulting interplay between the Jahn-Teller (JT)
effect and Hund's rule. The latter favors high-spin ground states
while the former splits the orbital degeneracy by a JT distortion of
the molecule, leading to a low-spin ground state. This in turn will influence the
tendency of hole-doped systems to become e.g. insulating or magnetic or 
superconducting.

This interplay can be observed in the electron-doped compounds, some of which
show metallic behavior and can become superconducting, such as 
A$_{3}$C$_{60}$, (A=K, Rb, Cs) \cite{A3C60-1,A3C60-2},
others are insulating like, e.g. Na$_{2}$C$_{60}$ \cite{Na2C60} or
A$_{4}$C$_{60}$  \cite{A4C60}.  Recent data on the charge transfer compounds
(AsF$_6$)$_2$C$_{60}$ and (SbF$_6$)C$_{60}$, which contain nominal
C$_{60}$$^{2 +}$ seem insulating and compatible with a magnetic ground state of 
the C$_{60}$$^{2 +}$ ions \cite{Datars}.

In this paper we report a theoretical study of the multiplet structures
of isolated C$_{60}^{\pm n}$ ions. Due to the open shell nature of the charged molecules, 
the multiplet energies cannot be obtained reliably by more standard methods such as 
ROHF ({\em Restricted Open-Shell Hartree-Fock}) or LDA ({\em Local Density Approximation })
calculations. 
We apply {\em ab initio} CAS SCF ({\em Complete Active Space Self-Consistent Field}) 
calculations, which are the appropriate generalization of the Hartree-Fock (HF) method
to orbitally degenerate systems \cite{cas}.

We compare our results with previous studies, which were based on the
semi-empirical calculations \cite{semi-korea,semi-okada}, model-SCF calculations
\cite{Nikolaev}, limited CI ({\em Configuration Interaction}) calculations \cite{Szabo,CI-Saito-mult} 
and on constrained LDA calculations \cite{LuedersEtAl:02}. 


\section{Model Hamiltonian}

As the starting point for describing the low energy physics of doped fullerenes 
one needs a model Hamiltonian which describes the interactions among the $t_{1u}$ 
electrons in case of electron-doping and or the $h_u$ holes in case of hole-doping, 
coupled to the vibrations of atoms in the molecule. 
Such a model was recently proposed in Ref. \cite{LuedersEtAl:02} and shall only be 
briefly introduced here.

The model Hamiltonian for a single molecule reads
\be
\hat{H} = \hat{H}_0 + \hat{H}_{\rm vib}  + 
\hat{H}_{\rm e-vib} + \hat{H}_{\rm  e-e},
\label{modelhamiltonian}
\ee
where 
\bea
\hat{H}_0     &=& \epsilon \, \sum_{\sigma m}  
\hat{c}^\dagger_{\sigma m} \hat{c}_{\sigma m}, \\
\hat{H}_{\rm vib} &=& \sum_{i\Lambda \mu} \frac{\hbar \omega_{i\Lambda}}{2} 
( \hat{P}_{i\Lambda \mu}^2 + \hat{Q}_{i\Lambda \mu}^2 ), \\
\hat{H}_{\rm e-vib} &=& 
\sum_{\sigma m m'} \sum_{r\,i\Lambda \mu}
\frac{g^r_{i\Lambda} \hbar \omega_{i\Lambda}}{2} 
C^{r \Lambda \mu}_{m m'} \, \hat{Q}_{i\Lambda \mu} \,\hat{c}^\dagger_{\sigma m}
\hat{c}_{\sigma m'}, \\
\hat{H}_{\rm  e-e} &=& 
\frac{1}{2} \sum_{\sigma, \sigma'} \sum_{{m m'}\atop{n n'}} 
w_{\sigma,\sigma'}(m,m';n,n') \, 
\hat{c}^\dagger_{\sigma m} \hat{c}^\dagger_{\sigma' m'} \, 
\hat{c}_{\sigma' n'} \hat{c}_{\sigma n},
\label{Coulomb-hamiltonian}
\eea

\noindent
are respectively the single-particle Hamiltonian, the molecular vibration energy 
in the harmonic approximation, the electron-vibration coupling in the
linear JT approximation \cite{Manini01,hbyh}, and the mutual 
Coulomb repulsion between the electrons.
The $\hat{c}^\dagger_{\sigma, m}$ denote the creation operators of either
a hole in the $h_u$ HOMO ({\em Highest Occupied Molecular Orbital})
or an electron in the $t_{1u}$ LUMO ({\em Lowest Occupied Molecular Orbital}), described by the
single-particle wave function $\varphi_{m\sigma}(\vr)$. 
The index $\sigma$ is the spin projection, $m$ and $n$ label the component
within the degenerate electronic HOMO/LUMO multiplet, and $i$ counts the
vibration modes of symmetry $\Lambda$ (2 $A_g$, 6 $G_g$ and 8 $H_h$ modes).
$C^{r \Lambda \mu}_{m n}$ are Clebsch-Gordan coefficients of the
icosahedral group, for coupling $h_u$ (holes) or $t_{1u}$ (electrons)
states to vibrations of symmetry $\Lambda$.
The index $r=$1,2 is a multiplicity label, relevant for $H_g$ modes only
\cite{Manini01,Butler81}.
$\hat{Q}_{i\Lambda \mu}$ and $\hat{P}_{i\Lambda \mu}$ are the molecular
vibration coordinates and conjugate momenta.

The Coulomb matrix elements are defined by:
\begin{equation}
\label{Coulomb-ints}
\fl w_{\sigma,\sigma'}(m,m';n,n') =
\Id{r} \! \Id{r'} \, 
\varphi^{*}_{m \sigma}(\vr) \, 
\varphi^{*}_{m'\sigma'}(\vr') \, 
u_{\sigma,\sigma'}(\vr,\vr') \, 
\varphi_{n\sigma}(\vr) \,
\varphi_{n'\sigma'}(\vr'), 
\end{equation}
where $u_{\sigma,\sigma'}(\vr,\vr')$ is a generally screened Coulomb interaction. 
In Ref. \cite{LuedersEtAl:02} it was demonstrated that this interaction can,
without loss of generality in the given subspace, be expressed in terms of a 
minimal number of physical parameters as: 
\be
\hat{H}_{\rm e-e} = \frac{1}{2} \sum_{r r' \Lambda} F^{r r' \Lambda} 
\left( \sum_\mu \hat{w}^{r \Lambda \mu} \, \hat{w}^{r' \Lambda \mu} \right)
- A \hat{n}, 
\label{He-ecombined}
\ee
where we defined the operators:
\be
\hat{w}^{r \Lambda \mu} = \sum_{\sigma} \sum_{m n} 
C^{r \Lambda \mu}_{m n} \, \hat{c}^\dagger_{\sigma m} \hat{c}_{\sigma n}.
\ee
The single-particle term $A$, which results from rearranging the field operators,
is a function of the Coulomb parameters defined below.

For electrons, the allowed Coulomb parameters, which generalize Slater's Coulomb parameters
for atoms \cite{Slater}, arise from the product $t_{1u} \otimes t_{1u} = A_g \oplus H_g$ and we
define thus $F_1 = F^{A_g}$ and $F_2 = F^{H_g}$. For holes, the
relevant product is $h_u \otimes h_u = A_g \oplus G_g \oplus 2 H_g$,
giving rise to five parameters $F_1 = F^{A_g}, F_2 = F^{G_g}, F_3 = F^{1,1,H_g},
F_4 = F^{2,2,H_g}$ and the cross-term $F_5 = F^{1,2,H_g}$.

Interaction parameters with a more transparent meaning can be further introduced by looking 
at the averaged energies of a multiplet of given charge $n$ and spin $S$. We find that
\be
\label{U_ave_model}
E_{\rm ave}(n) = E_0 + \epsilon n + U \frac{n (n-1)}{2}
\ee
and 
\be
E_{\rm ave}(n,S) - E_{\rm ave}(n,S-1) = 2 \tilde{J}S
\label{aveJ}
\ee
for both electrons and holes. To make contact with standard notation established  
historically, we rescale the exchange parameter for electrons as $J = 4\tilde{J}$/5, 
but keep $J = \tilde{J}$ for holes. 
It should be noted that here the on-site Coulomb repulsion parameter $U$ is defined 
with respect to the multiplet averaged energies, while traditionally the "Hubbard" 
$U$ (we call it $U_{0}$ in this work) is defined with respect to the ground-state 
energies of each charged system, i.e. 
\begin{equation}
U_0 = I(n)-A(n) = I(n)-I(n+1) = E_0(n+1) + E_0(n-1) - 2 E_0(n),
\label{U0}
\end{equation}
where $I(n)$ and $A(n)$ are respectively the ionization potential and the 
electron affinity of the n-charge ion.

The parameters introduced above can be expressed in terms of the
original Coulomb-parameters, as:
\bea
U_{t_{1u}} &=& \frac{1}{3} \left( F_1 - F_2 \right) \\
J_{t_{1u}} &=& 2 F_2 \\
A_{t_{1u}} &=& \frac{1}{2} U + 2 J 
\label{model-}
\eea
for electrons, and
\bea
U_{h_u} &=& \frac{F_1}{5} - \frac{4 \, F_2}{45} - \frac{F_3}{9} - \frac{F_4}{9}  \\
J_{h_u} &=& \frac{1}{6} F_2 +\frac{5}{24} (F_3 + F_4) \\
A_{h_u} &=& \frac{1}{2} U + \frac{8}{3} J  
\label{model+}
\eea
for holes. Analytic expressions for the eigenenergies of 
Hamiltonian \ref{modelhamiltonian} for holes were partly already reported
in Ref. \cite{LuedersEtAl:02}. For completeness, we provide the
eigenenergies for the $N$ electron states in Table~\ref{e-energies}.
Our scope in the rest of this paper will be the calculation of all these Coulomb 
parameters.

\begin{table}
\caption{Multiplet energies of the $N$ electron states $(t_{1u})^N$
relative to the average energy $\epsilon N + U N(N+1)/2$.
\label{e-energies}}
\begin{indented}
\item \begin{tabular}{ccr|ccr}
\br
N & ${}^{S (S+1)}\Lambda$ & $E_{\rm mult}$ & 
N & ${}^{S (S+1)}\Lambda$ & $E_{\rm mult}$ \\
\mr
2,4 & ${}^1 A_g$    & 4$J$ & 3   & ${}^2 T_{1u}$ &  2$J$ \\ 
    & ${}^1 H_g$    &   $J$ &     & ${}^2 H_u$    &  0    \\
    & ${}^3 T_{1g}$ & -$J$ &     & ${}^4 A_u$    & -3$J$ \\
\br
\end{tabular}
\end{indented}
\end{table}

\section{Computational details}

In order to find the multiplet splittings we performed CAS SCF
\cite{cas} calculations using the quantum chemistry package GAMESS
\cite{gam1}.  If not stated otherwise, all calculations presented here
have been performed using the 6-31 Gaussian basis set
[6s3sp1sp]/[1s2sp] \cite{bigbas}.

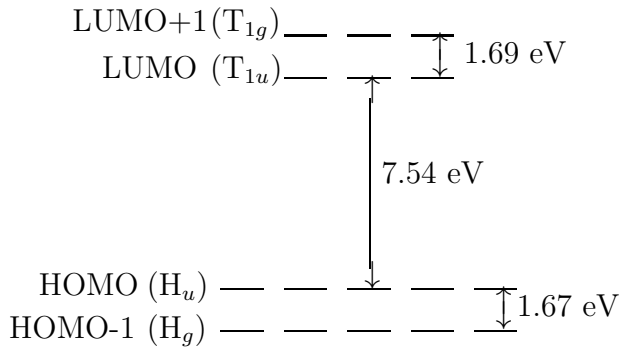
\begin{figure}
\caption{\label{HOMO-LUMO}The orbital degeneracies and splittings 
close to the HOMO-LUMO gap.}
\begin{center}
\unitlength0.8pt
\begin{picture}(400,185)(0,0)
\put(140,150){\line(1,0){20}}
\put(170,150){\line(1,0){20}}
\put(200,150){\line(1,0){20}}
\put(140,130){\line(1,0){20}}
\put(170,130){\line(1,0){20}}
\put(200,130){\line(1,0){20}}
\put(110,30){\line(1,0){20}}
\put(140,30){\line(1,0){20}}
\put(170,30){\line(1,0){20}}
\put(200,30){\line(1,0){20}}
\put(230,30){\line(1,0){20}}
\put(110,10){\line(1,0){20}}
\put(140,10){\line(1,0){20}}
\put(170,10){\line(1,0){20}}
\put(200,10){\line(1,0){20}}
\put(230,10){\line(1,0){20}}
\put( 75,7){(H$_{g}$)}
\put( 75,27){(H$_{u}$)}
\put(105,152){(T$_{1g}$)}
\put(105,130){(T$_{1u}$)}
\put(10,7){HOMO-1}
\put(24,27){HOMO}
\put(40,152){LUMO+1}
\put(54,130){LUMO}
\put(210,133){$\downarrow$}
\put(210,141){$\uparrow$}
\put(240,14){$\downarrow$}
\put(240,21){$\uparrow$}
\put(178,33){$\downarrow$}
\put(178,121){$\uparrow$}
\put(180.7,40){\line(0,1){80}}
\put(225,137){1.69 eV}
\put(186,80){7.54 eV}
\put(250,17){1.67 eV}
\end{picture}
\end{center}
\end{figure}

The energetic picture of the canonical orbitals,  shown in Figure~\ref{HOMO-LUMO},
motivates the choice of the active space.
For the positive ions, we consider only states
with holes delocalized in the 5-fold degenerate HOMO shell,
and for the negative ions, the states with
electrons in the 3-fold degenerate LUMO shell.
This is justified by the large HOMO-LUMO gap of the isolated C$_{60}$ 
molecule 7.54 eV.  The separation between the HOMO shell and the HOMO-1 
shell is 1.67 eV.  A similar gap of 1.69 eV is found between the LUMO 
and LUMO-1 shells.
We checked moreover that including the HOMO-1 and LUMO+1 shells respectively
did not change the results for one and two holes in the HOMO and one and
two electrons in the LUMO.  Thus, all results are obtained
with only one degenerate shell in the active space.
The natural orbitals were optimized with respect to the averaged
energy of all possible multiplet states regardless of the spin
multiplicity.
The geometry for the neutral C$_{60}$ was optimized with the RHF ({\em Restricted 
Hartree-Fock}) method with perfect icosahedral symmetry. The hexagon and pentagon
bond lengths obtained are 1.474~\AA $\;$ and 1.391~\AA $\;$ respectively. These
are not far from the experimental crystalline values obtained from the neutron powder
diffraction study (1.455~\AA $\;$ and 1.391~\AA $\;$ \cite{expbond1}), 
and $^{13}$C NMR measurement of C-C bonds (1.450~\AA $\;$ and 1.40~\AA $\;$ \cite{expbond2}). 
\footnote{ In order to further improve these results, one would need a bigger
basis-set and inclusion of dynamical correlations in the calculations. The
importance of these effects on the geometry and vibrational modes of
C$_{60}$ is discussed in Ref. \cite{Scuseria}. Performing dynamically
correlated calculations within the perturbation theory, for instance
on the MP2 level ({\em second order M\"{o}ller-Plesset}) \cite{MP21,MP22},
would be computationally very expensive for the systems studied here. MP2 
is a Rayleigh-Schr\"odinger perturbative method which uses the Hartree-Fock 
nonperturbed wave function and Hamiltonian as a zeroth order approximation. 
The perturbed functions depend on single-particle energies therefore it is 
a so-called "method with dynamical correlations". 
The semiempirical-approaches by the use of spectroscopic data for the estimation
of two-electron integrals and the LDA method by the parametrization on the QMC data 
({\em Quantum Monte Carlo}) also partially include the dynamical correlations.
To estimate the effect of dynamical correlations, we
report some results where we include them partially  for the $C_{60}^{1-}$ ion.}

Although the JT effect is crucial for the physics of C$_{60}$ ions
\cite{Manini01}, these changes of the geometry are relatively small and their influence
on the Coulomb integrals (see Eq. (\ref{Coulomb-ints})) should be fairly small. In the
present calculations we ignore therefore those distortions and fix the geometry
as frozen to that of neutral C$_{60}$.


\section{Results}

First we report the ground state energies for the C$_{60}$ ions
relative to the ground state of the neutral system.  They are shown in
Table \ref{Egs}, together with the ionization potentials and, if
known, the corresponding experimental values. We report also the Hubbard
$U$'s, calculated from the ground state energies.

\begin{table}[b]
\caption{\label{Egs}Ground state energies (in eV), ionization potentials and
Hubbard $U$'s ($U_{0}$, obtained from the ground state total energies
according to eq. (\ref{U0})) for the charged fullerene.}
\begin{indented}
\item[]
\begin{tabular}{r|rrcl}
\br
\mbox{Charge} & $E_{\rm GS}$ & $I$ & $I_{\rm exp}$ & $U_0$ \\
\mr
-6 &   43.157 &  -14.974  &                &     \\
-5 &   28.183 &  -12.009  &                &   2.97  \\
-4 &   16.174 &   -8.999  &                &   3.01 \\
-3 &    7.175 &   -5.415  &                &   3.58 \\
-2 &    1.760 &   -2.393  &                &   3.02 \\
-1 &   -0.633 &    0.633  & 2.65 $\pm$ 0.05  &   3.03 \\
 0 &    0.000 &    7.883  & 7.6 $\pm$ 0.2    &    \\
 1 &    7.883 &   10.925  & 11.46 $\pm$ 0.05 &   3.04 \\
 2 &   18.808 &   13.989  &                &   3.06 \\
 3 &   32.797 &   17.052  &                &   3.06 \\
 4 &   49.849 &   20.110  &                &   3.06 \\
 5 &   69.959 &   24.336  &                &   4.26 \\
 6 &   94.295 &   27.403  &                &   3.07 \\
 7 &  121.698 &   30.492  &                &   3.09 \\
 8 &  152.190 &   33.591  &                &   3.10 \\
 9 &  185.781 &   36.6728  &                &   3.08 \\
10 &  222.453 &            &                &        \\
\br
\end{tabular}
\end{indented}
\end{table}

The calculated ionization potential (IP) in the +1 charged case 10.93~eV 
is in a good agreement with the experimental value
11.46$\pm$0.05~eV \cite{IP+1}.  Also for the neutral molecule the
calculated IP at 7.88~eV is close to the measured 7.6$\pm$0.2~eV
\cite{IP0}.  Not surprisingly, the IP for the negative ion at 0.63~eV 
differs much from the experimental 2.65$\pm$0.05~eV \cite{IP-1}.  
The negatively charged ion is theoretically much more demanding since the
additional electron moves in a less binding potential, is very
delocalized and very sensitive to correlations with the other electrons.
Therefore the basis set and the relaxation
effects are more important in negative ions. 
To check the effects of the basis set and of additional correlations,
we also performed calculations with the minimal basis set of order 3-21 
([6s3p]/[2s1p]) \cite{MINI}, which also allowed for a ROMP2 ({\em Restricted 
Open Shell MP2}) calculation, with frozen 175 occupied orbitals and 10+1
correlated electrons. The restricted basis set reduced the IP of
C$_{60}^{-}$, as expected, to -0.03~eV. Inclusion of the dynamical
correlation increased the value from -0.03~eV to +0.43~eV. 
We checked that inclusion of 5-fold  degenerate orbitals on HOMO-1 level 
in the active space does not change  the IP for the C$_{60}^{-}$ ion by more than 0.06~eV. 
Also the inclusion of higher 3-fold degenerate  orbitals gives minor changes. 
The excitations from these orbitals contribute to the multiplets of negative ion with 
total weight smaller than 2$\%$ and do not change any of the splittings. 
The better agreement with experiment obtained by other authors
by means of semiempirical approaches
\cite{semi-korea,semi-alamos,semi-Morokuma} for C$_{60}^{-}$
could be partly due to the fact that the parameters used there are derived from spectroscopic
data. 
 C$_{60}^{2-}$ ions were observed in gas phase measurements
\cite{n2exp1,n2exp2} and found to be stable in semiempirical calculations
\cite{semi-korea,semi-alamos}. However the experimental results were questioned by other
group \cite{2-exp}.
In print of principle, our calculations are not ideally suited for negative ions. First of all, 
the weakly bound electron state is delicately sensitive to polarization and to correlations, not
very accurate in DFT. Secondly, the basis set becomes critical unless it is extended very much,
which is beyond our scopes. Earlier calculations by Razafinjanahary {\em et al.} \cite{Chermette}
are probably of better quality in that respect. 
A comparison to the previous calculations by means of DFT ({\em Density Functional Theory})
given in Table 4 of Ref.~\cite{Chermette} supports the importance of electronic
correlations. The IPs collected in that paper strongly depend on the exchange-correlation 
functional as one can see from values: 2.0~eV \cite{new1}, 1.9~eV \cite{Chermette}, 2.7~eV \cite{new1} 
and 2.8~eV \cite{new2} for C$_{60}^{-}$ and values: 
-1.3~eV \cite{new1}, -1.2~eV \cite{Chermette}, -0.4~eV \cite{new1} and -0.3~eV \cite{new2}
for C$_{60}^{2-}$ obtained respectively with functionals: $X\alpha$, VWN (Vosko-Wilk-Nusair), 
BH (Barth-Hedin) and PZ (Perdew-Zunger).
Our IP of C$_{60}^{3-}$ found to be -5.42~eV should be compared with value -4.4~eV obtained
with the VWN functional from the modified MT ({\em Muffin-Tin Approximation}) calculations 
by Razafinjanahary {\em et al.} \cite{Chermette}. \footnote{The question was raised in 
Ref. \cite{Chermette} about possible electron states trapped inside the C$_{60}$ molecule. 
If such states existed, a calculation like the present where all basis functions are carbon-centered, 
and none is molecule-centered, could very well miss such a trapped state. However, the trapped 
state would not be missed in a plane-wave (PW) calculation. Direct comparison of our $t_{1u}$ LUMO
and $t_{2g}$ LUMO+1 derived states with those of the PW calculations of C$_{60}$ 
in Ref. \cite{PWcalc2, PWcalc1} shows perfect agreement, with no additional low laying states in 
the PW case that could be taken as trapped states (states with the spherical symmetry).} 

The bare molecular Hubbard $U$'s ($U_{0}$) are all of the order of 
3.01-3.10~eV for both  electrons and holes (see Table~2). 
Exceptions are the 3-fold negative and the 5-fold positive
systems. Here the $U$'s are considerably larger. 
The parameters $U_{0}$ obtained for the charge $n$ were derived from the ground state 
energies of molecules with charge $n-1$ and $n+1$. Due to the particle-hole symmetry, the
multiplet splittings of ions with charge one more and one less than the charge at half-filling
are identical (the energy gaps on right and left from half-filling are equal).
This is not the case of any other charge, where the "neighboring" multiplet splittings
are different and the ground energies of the charge multiplets which are more rich in 
the number of states, are lowered in comparison to the ground energies of multiplets 
with smaller number of states. The aforementioned asymmetry leads to smaller values of 
$U_{0}$ away from exact half filling. 

\begin{table}
\caption{\label{multiplet-234}
Multiplet energies (in Merv) for 2,3 and 4 electrons in the $(t_{1u})$ LUMPY,
and energies reconstructed from the model (eq. (11)-(13)) 
using the charge-averaged parameters eq. (9,10).}
\begin{indented}
\item[] \begin{tabular}{crrr|crr}
 \br
$\lambda$   & N=2 & N=4 & \mbox{Model} & $\lambda$ & N=3 & \mbox{Model} \\
 \mr
${}^3 T_{1g}$  &  -114.1 &  -112.7 & -113.5 & ${}^4 A_u$     & -340.7 & -340.4  \\
${}^1 H_g$     &   114.1 &   112.7 &  113.5 & ${}^2 H_u$     &    0.0 & 0.0 \\
${}^1 A_g$     &   456.6 &   450.9 &  453.9 & ${}^2 T_{1u}$  &  227.1 & 226.9    \\
 \br
 \end{tabular}
\end{indented}
\end{table}

\begin{table}
\caption{\label{multiplet+2378}
Multiplet energies (in meV) for 2,3,7 and 8 holes in the $(h_u)$ HOMO,
and energies reconstructed from the model (eq. (14)-(16))
using the charge-averaged parameters eq. (9,10).}
 \begin{indented}
 \item[] \begin{tabular}{crrr|crrr}
 \br
 $\lambda$  & N=2 & N=8 & \mbox{Model} &                 $\lambda$  & N=3 & N=7 & \mbox{Model} \\ 
 \mr
${}^3 T_{b g}$      &  -134.8 &  -125.2 &  -133.8 &      ${}^4 T_{bu}$              &   -388.9 &  -382.3 & -390.3 \\
${}^3 T_{a g}$      &  -127.9 &  -139.7 &  -128.7 &      ${}^4 T_{au}$              &   -385.8 &  -394.8 & -385.2 \\
${}^3 G_g$          &  -122.4 &  -122.5 &  -123.8 &      ${}^4 G_u$                 &   -378.8 &  -382.0 & -380.2 \\
${}^1 G_g$          &    52.0 &    63.6 &    57.7 &      ${}^2 T_{au}$              &   -113.6 &   -99.2 & -109.1 \\
${}^1 H_g$          &    57.4 &    67.1 &    62.2 &      ${}^2 T_{bu}$              &   -110.5 &  -107.8 & -104.0 \\
${}^1 H_g$          &   463.3 &   448.4 &   455.9 &      ${}^2 H_u$                 &   -105.6 &   -95.3 & -101.3 \\
${}^1 A_g$          &  1022.3 &  1029.7 &  1026.0 &      ${}^2 H_u$                 &   -102.2 &   -91.3 &  -97.4 \\
 & & & &                                                 ${}^2 G_u$                 &    191.2 &   189.1 &  190.1 \\
 & & & &                                                 ${}^2 T_{1u},{}^2 T_{2u}$  &    199.7 &   196.0 &  197.7 \\
 & & & &                                                 ${}^2 H_u$                 &    398.1 &   386.9 &  392.1 \\
 & & & &                                                 ${}^2 G_u$                 &    399.1 &   388.1 &  393.1 \\
 & & & &                                                 ${}^2 H_u$                 &    767.9 &   772.1 &  768.6 \\
 \br
 \end{tabular}
 \end{indented}
\end{table}

\begin{table}
\caption{\label{multiplet+456}
Multiplet energies (in meV) for 2,3,7 and 8 holes in the $(h_u)$ HOMO,
and energies reconstructed from the model (eq. (14)-(16)) 
using the charge-averaged parameters eq. (9,10).}
 \begin{indented}
 \item[]\begin{tabular}{crrr|crr}
 \br
 $\lambda$  & N=4 & N=6 & \mbox{Model} & $\lambda$ & N=5 & \mbox{Model} \\
 \mr
${}^5 H_g$      &  -768.3 &  -770.1  & -769.5 &   ${}^6 A_u$    &  -1282.7 & -1282.6 \\
${}^3 G_g$      &  -395.7 &  -389.8  & -393.1 &   ${}^4 H_u$    &   -712.6 &  -712.4 \\ 
${}^3 H_g$      &  -394.7 &  -388.9  & -392.2 &   ${}^2 H_u$    &   -331.3 &  -331.1 \\ 
${}^1 G_g$      &  -209.5 &  -199.7  & -204.9 &   ${}^2 G_u$    &   -330.8 &  -330.6 \\ 
${}^1 A_g$      &  -205.5 &  -196.2  & -201.2 &   ${}^4 H_u$    &   -318.7 &  -318.7 \\ 
${}^3 G_g$      &  -197.0 &  -195.2  & -196.2 &   ${}^4 G_u$    &   -314.2 &  -314.2 \\ 
${}^3 T_{b g}$  &  -194.9 &  -196.3  & -195.6 &   ${}^2 A_u$    &   -139.1 &  -139.0 \\ 
${}^3 T_{a g}$  &  -193.9 &  -189.7  & -192.1 &   ${}^4 G_u$    &   -132.8 &  -132.8 \\ 
${}^1 H_g$      &    86.2 &    88.9  &   87.7 &   ${}^2 G_u$    &   -131.5 &  -131.4 \\ 
${}^1 T_{a g}$  &    92.0 &    91.0  &   91.7 &   ${}^4 T_{au}$ &   -127.7 &  -127.8 \\ 
${}^1 H_g$      &    93.8 &    97.9  &   95.5 &   ${}^4 T_{1u}$ &   -122.7 &  -122.7 \\ 
${}^1 T_{b g}$  &    93.5 &    97.2  &   96.7 &   ${}^2 H_u$    &    143.6 &   143.8 \\ 
${}^3 T_{a g}$  &   103.5 &   100.6  &  100.6 &   ${}^2 H_u$    &    156.1 &   156.1 \\ 
${}^3 H_g$      &   103.3 &    98.3  &  101.9 &   ${}^2 T_{bu}$ &    158.9 &   158.9 \\ 
${}^3 T_{b g}$  &   104.0 &   100.6  &  102.4 &   ${}^2 H_u$    &    159.0 &   159.0 \\ 
${}^3 H_g$      &   105.9 &   100.6  &  103.4 &   ${}^2 T_{au}$ &    162.9 &   162.8 \\ 
${}^3 G_g$      &   382.6 &   382.0  &  383.4 &   ${}^2 T_{au}$ &    164.3 &   164.4 \\ 
${}^1 A_g$      &   384.3 &   383.8  &  383.4 &   ${}^2 H_u$    &    166.9 &   166.9 \\ 
${}^3 T_{a g}$  &   385.1 &   384.7  &  385.0 &   ${}^2 T_{bu}$ &    167.9 &   167.9 \\ 
${}^3 T_{b g}$  &   385.6 &   387.9  &  386.7 &   ${}^2 T_{bu}$ &    443.9 &   443.7 \\ 
${}^1 G_g$      &   392.3 &   388.8  &  390.9 &   ${}^2 T_{au}$ &    453.4 &   453.4 \\ 
${}^1 H_g$      &   564.3 &   569.9  &  567.2 &   ${}^2 G_u$    &    455.7 &   455.6 \\ 
${}^1 G_g$      &   578.4 &   581.1  &  580.2 &   ${}^2 G_u$    &    460.1 &   459.9 \\ 
${}^1 H_g$      &   590.4 &   583.1  &  587.2 &   ${}^2 A_u$    &    463.9 &   463.6 \\ 
${}^1 G_g$      &   591.3 &   585.2  &  588.3 &   ${}^2 G_u$    &    646.8 &   646.7 \\ 
${}^1 H_g$      &   970.3 &   966.9  &  969.0 &   ${}^2 H_u$    &    651.3 &   651.1 \\
${}^1 A_g$      &  1536.7 &  1540.7  & 1539.2 &   ${}^2 H_u$    &   1026.3 &  1026.2 \\

 \br
 \end{tabular}
\end{indented}
\end{table}

The key results of the above described CAS SCF calculations are
the multiplet energies of the fullerene ions. They are
reported in Table~\ref{multiplet-234} for electrons
and Tables~\ref{multiplet+2378} and \ref{multiplet+456} for holes.
Since GAMESS does not support symmetries as high as icosahedral, the assignment of
the symmetry labels to the states in CAS SCF calculations was done indirectly, 
using the degeneracies of the states. In this way, the three-fold degenerate $T_1$ and $T_2$ 
representations cannot be distinguished.  However, due to the analytical 
results of the model Hamiltonian, reported in Ref. \cite{LuedersEtAl:02}, 
they can be assigned once a certain sign of the parameter $F_5$ is chosen. 
Hence, in these cases we label the states by $T_{a/b}$ where $a/b$ represents 
$1$ or $2$ respectively. In the negative ions, no such ambiguity arises.

We find that the spread of the multiplets is about 0.6~eV for all
negatively charged states, 1.2~eV for 2, 3, 7 and 8 holes, and 2.3~eV 
for 4-6 holes. Comparing the spectra for $N$ and 6-$N$ electrons, or $N$
and 10-$N$ holes respectively, one finds that the particle-hole symmetry in
the occupation of the degenerate orbitals is approximately preserved
in both cases. \footnote{Clearly however, this approximate symmetry is a result
of out single-orbital approximation, and would disappear once all other levels would
be included. In this sense, it is mostly of academic significance.} 

The semi-empirical calculations for multiplets of
C$_{60}^{n-}$ ions and for
C$_{60}^{+1}$ and C$_{60}^{+2}$ ions \cite{semi-korea} give larger
splittings than ours probably due to unequal occupation of degenerate
orbitals.  
The limited CI calculations with HF reference orbitals  do not take care of the right 
degeneracies due to the splitting of $t_{1u}$ shell when it is partially occupied
by the electrons. The discussion of this fact is given in Ref. \cite{CI-Saito-mult}. 
The CAS SCF calculations are free of symmetry related problems. 
Also, excessively large splittings were obtained with the restricted configuration interaction 
constructed using the INDO ({\em Intermediate Neglect of the 
Diatomic Overlap}, for the method see \cite{INDO}) orbitals \cite{semi-okada}.

Recently a complete analysis of the multiplets of all C$_{60}$ ions
was presented \cite{Nikolaev}, where the multiplet energies were
evaluated from a multipole expansion of the Coulomb interaction. The
Coulomb integrals were calculated with respect to symmetry-adapted
model wave functions.  The obtained multiplet spectra are in good 
agreement with our results, apart from a larger $T_1$/$T_2$ splitting, and a
change in the level ordering.  We will comment on this point
later, when discussing the Coulomb parameters extracted from the
present results.

We observe a degeneracy between the ${}^2 T_{1u}$ and the ${}^2 T_{2u}$ states of the three 
hole multiplets. This apparently accidental degeneracy is indeed well studied in the literature
\cite{Oliva:97,LoJudd:99,Plakhutin:00} and attributed to the permutational symmetry.


\section{Determination of the model parameters}

According to Eq.~(\ref{U_ave_model}), the multiplet-averaged energies of 
the model are described by the parabola:
\be
E_{\rm ave}(n) = E_0 + (\epsilon - \frac{U}{2}) n + U \frac{n^2}{2},
\ee 
where $n$ denotes the number of electrons or holes respectively.

\begin{figure}
\caption{\label{fig-fit}
Average multiplet energies and the fitted parabola for
the $(t_{1u})^N$ states for electrons and the $(h_{u})^N$ states
for holes.}
\vspace{0.5cm}
\centerline{
\includegraphics[scale=0.4,angle=-90]{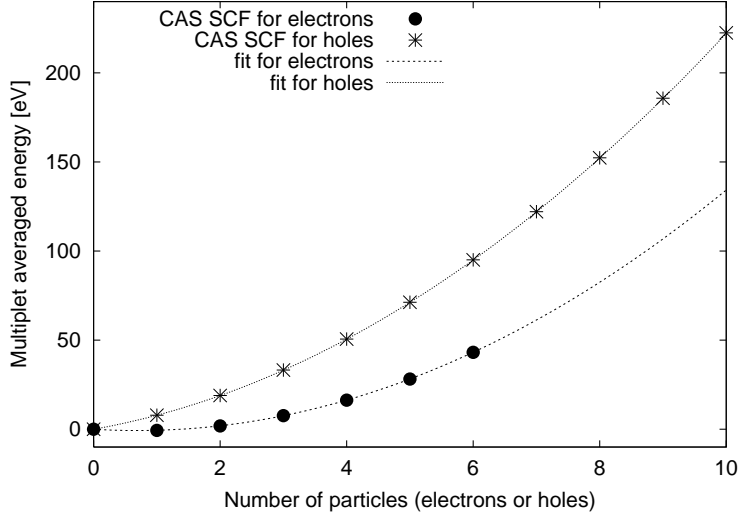}}
\end{figure}

Figure~\ref{fig-fit} displays the calculated average 
energies, together with the fitted parabolas:
\bea
E_{\rm ave}^{e^-}(n) &=& -61393.320 - 2.145 n + 1.557 n^2, \\
E_{\rm ave}^{h^+}(n) &=& -61393.268 + 6.245 n + 1.599 n^2.
\eea
From these parabolas the average Coulomb repulsion $U_{\rm ave}$ and the 
single particle energies $\epsilon$ are obtained. These are displayed 
in Table~\ref{U-fit}. 

\begin{table}
\caption{\label{U-fit} Coulomb parameters end Koopmans single particle energies
for electrons and holes (in eV).} 
\begin{indented}
\item[]\begin{tabular}{lrrr}
\br
 & $U_{\rm ave}$ & $\epsilon$ & $\epsilon_{\rm RHF}$ \\
\mr
electrons & 3.1141 & -0.5881 & -0.4898 \\
holes     & 3.1928 & -7.8477 & -8.0270 \\
\br
\end{tabular}
\end{indented}
\end{table}

This Coulomb repulsion $U_{\rm ave}$, of about 3 eV for both electrons and
holes, is in good agreement with previous calculations
\cite{Antropov:92, LuedersEtAl:02}.
The single-particle energies, extracted from the parabolas, are very
close to the single-particle levels of the neutral system. It should
be noted that this is not an obvious finding, since the single-particle energies, 
extracted from the fit, derive from a  calculation of the charged molecules. 
In the model Hamiltonian, the first ionization potential is equal exactly 
to the single particle energy of the holes. 
In fact, Koopmans theorem states that the HOMO level gives a 
good approximation for the ionization potential.
In general Koopmans theorem does not perform very well for the
electron affinity, due to a bad description of the virtual states by
means of the HF method (see \cite{Szabo} p. 127-128). 
Calculations involving charged states, in which the LUMO shell of 
the neutral molecule is actually occupied,
should give much better values. Here we find, however, that the single
particle energy of the electrons, determined from the fit, which
represents the electron affinity of the model, is very close to the
LUMO level obtained by means of the RHF method.  

\begin{table}
\caption{\label{J-e} Exchange splitting $J$ of the electron states $(t_{1u})^N$, 
for each charge $N$, as well as it's averaged value (ave).
The values labeled by limited CI are extracted from  \cite{CI-Saito-mult}, 
INDO+CI from \cite{semi-okada}, "direct" refers to  Nikolaev {\it et.al.}~\cite{Nikolaev}, 
LDA to L{\"u}ders {\it et al.}~\cite{LuedersEtAl:02}, MNDO+CI to  \cite{semi-alamos}.
All parameters are given in meV.} 
\begin{indented}
\item[]\begin{tabular}{ccccc}
\br
Method & N=2 & 3 & 4 & ave   \\
\mr
CAS SCF & 114.1 & 113.6 & 112.7 & 113.5  \\[0.1cm]
limited CI & 29-146 & 8-388 & 8.5-380 & 15-305  \\[0.1cm] 
INDO+CI & 120 & 185 & - &  153   \\[0.1cm]
"direct" & - & - & - & 99, 95 \\[0.1cm]
LDA    & - & - & - &  32 \\[0.1cm]
MNDO+CI & 50 & - & - & 50 \\ 
\br
\end{tabular}
\end{indented}
\label{Authors}
\end{table}

Next, for the $N$ electron states $(t_{1u})^N$, only the exchange parameter $J$
remains to be determined. Knowing the analytical eigenenergies
(Table~\ref{e-energies}), it can be directly extracted from the
calculated multiplets given in Table~\ref{multiplet-234}. The
exchange constants $J$s, obtained in this way, are given in
Table~\ref{J-e}.  We compare them to other results by means of the LDA 
method, and to model calculations ("direct" in Table~\ref{J-e}) and
to the limited CI method, and finally to the small configuration 
interaction performed on the INDO orbitals.  
Our results are close to those obtained by the model 
calculations~\cite{Nikolaev}, although slightly larger. 
By contrast the values of Hund's rule exchange $J$ obtained by LDA is 
a factor of 3 smaller \cite{LuedersEtAl:02}.
The reason for the large discrepancy with the LDA calculations is
not clear. \footnote{This could be due to correlation effects included in the
effective Kohn-Sham potential or due to the symmetry or self-interaction problems
present in LDA calculations.} 

The $J$ parameters shown in Table~\ref{Authors} for limited CI are extracted according 
to Table~\ref{e-energies} from the multiplet splittings of either the lowest states 
or of the whole multiplet within the $t_{1u}$ shell. 
These CI parameters depend strongly on the charge and the
choice of states used for the derivation. This is due to the fact that the configuration
interaction approach does not treat correctly the symmetry of
$t_{1u}$ shell, removing the 3-fold degeneracy when the shell is partially occupied
by the electrons \cite{CI-Saito-mult}. CAS SCF calculations give correct degeneracies and
the obtained $J$ parameters are independent on the charge and the choice of states 
(due to the multiplet structure showed in Table~\ref{e-energies}). 
The fact that limited CI calculations underestimate splittings of the lowest states and 
overestimate splitting of higher states is not only due to the
broken symmetry, but also due to the use of HF single-particle orbitals optimized for the
ground state. Contrary to that, in our CAS SCF calculations we optimized the 
molecular orbitals in such a way that the total energy averaged over all states within 
the multiplet was minimal.

Other authors, by means of the MNDO+CI method ({\em Modified Neglect of Diatomic Overlap},
 for the method see \cite{MNDO}), derived $J$ = 30 meV for the neutral C$_{60}$ 
molecule and J=50 meV from the C$_{60}^{-2}$ triplet-singlet splitting \cite{semi-alamos}. 
The {\em ab initio} SCF calculations point to a $J$ value of order 100 meV \cite{Chang}.
The parameters obtained from the MNDO+CI  method are smaller than the  limited CI results
probably due to the use of semi-empirical orbitals (direct fit on spectroscopic data) 
for the CI construction.

\begin{table}
\caption{\label{symm_ave}
The Coulomb multiplets for C$_{60}^{N+}$, averaged over states of the same
representation, as a function of the e-e
parameters.
The model Hamiltonian (\ref{modelhamiltonian}) obeys particle-hole
symmetry: therefore the multiplet energies for $N>5$ holes equal those for
$(10-N)$ holes.
The non particle-hole symmetric contribution $\left[\epsilon N + U
N(N-1)/2\right]$ is left out in this table.
}
\begin{tabular}{lll}
\br
Ion & \mbox{State symmetry}    &  $E_{\rm mult, ave}(N,\Lambda,S)$ \\
\mr
$N$=2 & ${}^1 A_g$  &  $\frac{8}{9}  F_2 + \frac{10}{9}  F_3 + \frac{10}{9}  F_4$ \\
&  ${}^1 G_g$    &  $\frac{1}{18} F_2 - \frac{5}{36}  F_3 + \frac{25}{36} F_4$ \\
&  ${}^1 H_g  \ [\times 2]$ &  $\frac{2}{9}  F_2 + \frac{13}{36} F_3 + \frac{1}{36}  F_4$ \\
& ${}^3 T_{1g}$  & $-\frac{4}{9}  F_2 - \frac{5}{36}  F_3 + \frac{7}{36}  F_4 + \frac{\sqrt{5}}{2} F_5$ \\
& ${}^3 T_{2g}$  & $-\frac{4}{9}  F_2 - \frac{5}{36}  F_3 + \frac{7}{36}  F_4 - \frac{\sqrt{5}}{2} F_5$ \\
& ${}^3 G_g$  &  $\frac{7}{18} F_2 - \frac{5}{36}  F_3 - \frac{23}{36} F_4$ \\
\mr
$N$=3 & ${}^2 T_{1u}\ [\times 2]$ & $-\frac{1}{12} F_2 - \frac{1}{24} F_3 + \frac{7}{24}  F_4 - \frac{\sqrt{5}}{4} F_5$ \\
& ${}^2 T_{2u}\ [\times 2]$ & $-\frac{1}{12} F_2 - \frac{1}{24} F_3 + \frac{7}{24}  F_4 + \frac{\sqrt{5}}{4} F_5$ \\
& ${}^2 G_u   \ [\times 2]$ & $-\frac{1}{3}  F_2 + \frac{7}{12} F_3 + \frac{1}{12}  F_4$ \\
& ${}^2 H_u   \ [\times 4]$ &  $\frac{5}{12} F_2 + \frac{5}{24} F_3 + \frac{5}{24}  F_4$ \\
& ${}^4 T_{1u} $   & $-\frac{2}{3}  F_2 - \frac{5}{12} F_3 - \frac{1}{12}  F_4 + \frac{\sqrt{5}}{2} F_5$ \\
& ${}^4 T_{2u}$  & $-\frac{2}{3}  F_2 - \frac{5}{12} F_3 - \frac{1}{12}  F_4 - \frac{\sqrt{5}}{2} F_5$ \\
& ${}^4 G_u$  &  $\frac{1}{6}  F_2 - \frac{5}{12} F_3 - \frac{11}{12} F_4$ \\
\mr
$N$=4 & ${}^1 A_g   \ [\times 3]$ &  $\frac{5}{9}   F_2 +\frac{1}{2}   F_3 +\frac{17}{18} F_4$ \\
& ${}^1 T_{1g}$  &  $\frac{1}{2}   F_2 -\frac{1}{12}  F_3 +\frac{1}{4}   F_4 - \frac{\sqrt{5}}{2} F_5$ \\
& ${}^1 T_{2g}$  &  $\frac{1}{2}   F_2 -\frac{1}{12}  F_3 +\frac{1}{4}   F_4 + \frac{\sqrt{5}}{2} F_5$ \\
& ${}^1 G_g   \ [\times 4]$ &  $\frac{13}{24} F_2 +\frac{17}{48} F_3 +\frac{7}{48}  F_4$ \\
& ${}^1 H_g   \ [\times 5]$ &  $\frac{2}{15}  F_2 +\frac{17}{30} F_3 +\frac{17}{30} F_4$ \\
& ${}^3 T_{1g}\ [\times 3]$ &  $\frac{1}{18}  F_2 +\frac{1}{6}   F_3 -\frac{1}{18}  F_4$ \\
& ${}^3 T_{2g}\ [\times 3]$ &  $\frac{1}{18}  F_2 +\frac{1}{6}   F_3 -\frac{1}{18}  F_4$ \\
& ${}^3 G_g   \ [\times 3]$ &  $\frac{1}{18}  F_2 -\frac{1}{4}   F_3 +\frac{13}{36} F_4$ \\
& ${}^3 H_g   \ [\times 3]$ & $-\frac{1}{9}   F_2                    -\frac{2}{9}   F_4$ \\
& ${}^5 H_g $  & $-\frac{2}{3}   F_2 -\frac{5}{6}   F_3 -\frac{5}{6}   F_4$ \\
\mr
$N$=5 & ${}^2 A_u   \ [\times 2]$ & $-\frac5{18}    F_2 +\frac{13}{36} F_3 +\frac1{36}    F_4$ \\ 
& ${}^2 T_{1u}\ [\times 3]$ &  $\frac{4}{9}   F_2 +\frac{5}{18}  F_3 +\frac{1}{48}  F_4 + \frac{\sqrt{5}}{3} F_5$ \\
& ${}^2 T_{2u}\ [\times 3]$ &  $\frac{4}{9}   F_2 +\frac{5}{18}  F_3 +\frac{1}{48}  F_4 - \frac{\sqrt{5}}{3} F_5$ \\
& ${}^2 G_u   \ [\times 5]$ &  $\frac{7}{45}  F_2 +\frac{14}{45} F_3 +\frac{2}{45}  F_4$ \\
& ${}^2 H_u   \ [\times 7]$ &  $\frac{11}{63} F_2 +\frac{16}{63} F_3 +\frac{34}{63} F_4$ \\
& ${}^4 T_{1u}$   &  $\frac{2}{9}   F_2 -\frac{5}{36}  F_3 -\frac{17}{36} F_4 - \frac{\sqrt{5}}{2} F_5$ \\
& ${}^4 T_{2u}$   &  $\frac{2}{9}   F_2 -\frac{5}{36}  F_3 -\frac{17}{36} F_4 + \frac{\sqrt{5}}{2} F_5$ \\
& ${}^4 G_u   \ [\times 2]$ & $-\frac{4}{9}   F_2 -\frac{5}{36}  F_3 -\frac{11}{36} F_4$ \\
& ${}^4 H_u   \ [\times 2]$ & $-\frac{4}{9}   F_2 -\frac{23}{36} F_3 -\frac{11}{36} F_4$ \\
& ${}^6 A_u $  & $-\frac{10}{9}  F_2 -\frac{25}{18} F_3 -\frac{25}{18} F_4$ \\
\br
\end{tabular}
\renewcommand{\arraystretch}{1.0}
\end{table}

For the positively charged states, the situation is both more complicated,
and less explored.
Due to the uncertainty of the $T_{1}$, $T_{2}$ assignment and
the multiple occurrence of states of the same symmetry but different
energies, the parameters cannot be fitted directly to the multiplet energies.
We therefore refer to the average energies for a given symmetry, shown
in Table~\ref{symm_ave}.
As can be seen from this table, the averaging over 
the $T_1$ and the $T_2$ states depends only
on $F_2, F_3$ and $F_4$, while $F_5$ determines
$T_1$ - $T_2$ splitting only.
The parameters $F_2$ - $F_4$ can be determined completely, while $F_5$
can only be found up to a sign. As it turns out, $F_{5}$ is however small
enough to make this uncertainty irrelevant.

We fit the multiplet energies for each charge separately, allowing for 
charge-dependent model parameters, which are given in Table~\ref{params-h}.
\begin{table*}
\caption{\label{params-h}The model parameters for the positively 
charged states, determined from the multiplet spectra. We also give a 
comparison to the LDA calculations of Ref. \cite{LuedersEtAl:02} 
and the model (direct) calculations of Ref. \cite{Nikolaev}. 
All parameters are given in meV.}
\begin{indented}
\item[] \begin{tabular}{c|ccccccc|c|cc|c}
\br
& \multicolumn{8}{c|}{\mbox{CAS SCF}} & \multicolumn{2}{c|}{\mbox{direct}} & \mbox{LDA} \\
& $N$=2 & 3 & 4 & 5 & 6 & 7 & 8 & \mbox{ave} & II & III \\
\mr
 $F_2$  & 188.8 & 190.3 & 191.9 & 193.5 & 195.1 & 196.6 & 198.2 & 193.5   & 193 & 185 & 105 \\
 $F_3$  & 591.1 & 589.0 & 586.1 & 584.3 & 581.3 & 580.2 & 577.1 & 584.2   & 640 & 625 & 155 \\
 $F_4$  & 178.0 & 180.1 & 182.3 & 184.4 & 186.8 & 188.7 & 191.1 & 184.5   & 180 & 172 &  47 \\
$|F_5|$ &   3.1 &   1.4 &   0.7 &   2.2 &   4.3 &   5.6 &   7.6 & 2.3$^*$ &  13 &  13 &   0 \\
\mr
  $J$   & 191.7 & 191.9 & 192.1 & 192.4 & 192.5 & 193.0 & 193.1 & 192.4   & 203 & 197 &  60 \\
\br
\end{tabular}
\end{indented}
\end{table*}
The parameters $F_2,F_3$ and $F_4$ show a weak monotonic dependence on charge. 
This is in contrast to the result of Ref. \cite{Nikolaev}
whose parameters are basically determined at $N$=0, and then used for all $N$.
From this perspective it becomes clearer why the $T_{1u}/T_{2u}$ splittings
of Ref. \cite{Nikolaev} are much larger than ours.

The substantial difference of our CAS SCF results with the LDA parameters
of Ref. \cite{LuedersEtAl:02} for hole-doped ions, also shown in 
Table~\ref{params-h}, is similar to the discrepancy noted in the electron-doped case.  


\section{Summary}

The central aim of this work is a calculation of all Coulomb parameters
and multiplet energies for C$_{60}^{\pm n}$ ions including a proper 
treatment of the orbital degeneracies, within the icosahedral symmetry of C$_{60}$. 
This has been done within the CAS SCF framework.  

The model parameters for positively charged C$_{60}$ ions are 
substantially larger than the LDA results as in the
negatively charged molecules. Our results on the other hand, are rather closer to
model studies by Nikolaev \cite{Nikolaev}. A possible reason is that
both  approaches do not take into account the dynamical correlations, which
are very important, as was shown in Ref. \cite{Scuseria}. In the LDA and
MNDO methods, where some dynamical correlations are included, the multiplet
splittings are much reduced.
The limited CI method led in previous calculations \cite{CI-Saito-mult} to much smaller or much larger
$J$ values depending on the choice of states and on the charge case used for the
derivation of parameters. This could be due to the artificially broken symmetries  
when the electrons occupy partially the molecular orbitals. The parameters
obtained from the CAS SCF method, involving a correct treatment of symmetry,
do not depend on the charge or choice of states within the multiplet used for
the derivation.

The mean Coulomb parameters, $U$, obtained in this work from averaged multiplets, 
are for positively charged fullerenes about 3.2~eV. 
Similarly for the negative ions the obtained $U$ parameters are  about 3.1~eV.
The charge specific Hubbard $U_{0}$ are generally close to the average, except at half
filling where $U_{0}$ is not influenced by $J$. Here $U_{0}$=3.58 for C$_{60}^{3-}$ 
and $U_{0}$=4.26 for C$_{60}^{5+}$.

The exchange parameters, $J$, for the negative ions are somewhat basis set
dependent due to the localization effects discussed earlier, and their
values are of about 114~meV for 6-31 basis set and 146~meV for the
minimal basis set \cite{MINI}.  
The more extended basis set contains more expanded Gaussians, 
while the smaller basis set constraining the electrons to a smaller volume, 
is likely to increase the exchange parameter $J$.  
The exchange parameters for hole-doped
molecules are less basis set dependent: they are of order 192~meV
for the 6-31 basis set and 208~meV for the minimal basis set. 
These values probably still represent an approximation in excess and it seems
possible that a further slight decrease could be found with substantially 
larger basis sets. \footnote{The quality of Gaussian basis set type is more dependent 
on the well chosen exponents than on their number. There is  no possibility for 
the saturation of a basis set similar to that can obtained with the plane-wave methods.}

Experimentally, the multiplet structure of C$_{60}$ ions in vacuum is still, as far as we know,
inaccessible. In particular probing the magnetic state of ions in a beam appears to be difficult.
Probing charged fullerene ions embedded in a rare gas matrix would seem a better possibility,
that could be explored with the purpose identifying the ground states properties for the various
charged species, possibly with EPR \footnote{We owe this suggestion to G.~Scoles.}. 
The Coulomb couplings of $C_{60}$ ions play an important role in the solid state compounds. 
In particular in alkali fullerides the $C_{60}$
molecule is negatively ionized while in Lewis acid fullerene salts \cite{Panich1,Panich2} 
it should be positively ionized. NMR longitudinal relaxation time data of basically all alkali 
fullerides indicate a "spin gap", separating the low spin $C_{60}$ ion ground state 
from high spin excitations \cite{Alloul}. These spin gaps are typically 100-150 meV, 
compatible with a value of $J$ of e.g. 50-70 meV. Comparison with our calculated value of 113 meV 
indicates an overestimate. While that may be due to a genuine solid
state effect, we cannot rule out the possibility that the agreement could be improved by
further extending the used basis set.

Finally, our predictions of a large $J$ for positive fullerene ions, 
which also imply that they should be 
with great probability magnetic \cite{LuedersEtAl:02} should be directly accessible by further 
experiments on the acceptor salts \cite{Panich1,Panich2} as well as in the recently discovered
$C_{60}^{+2}$ and $C_{60}^{+3}$ in solution \cite{Paolucci}.

\section{Acknowledgments}

We would like to thank Nicola Manini for a first reading and useful comments,
and acknowledge discussions with J.~G.~Soto Mercado and G.~E.~Santoro. 
We are also grateful to M. Ricco for discussions, and to F. Paolucci for
informing us of his new results on $C_{60}^{+2}$ and $C_{60}^{+3}$.
This work was supported by the European Union, contracts
ERBFMRXCT970155 (TMR FULPROP), covering in particular the 
postdoctoral work of M. L{\"u}ders,  and HPRI-CT-1999-00048 (MINOS) 
for computing time at the CINECA supercomputing center and a fellowship (ML).
Research in SISSA was also supported through MIUR FIRB RBAU0178R004, FIRB RBAU01LX5H,
and COFIN 2003-028141-007.

\end{document}